\documentclass[journal = jctcce, manuscript = article]{achemso}

\usepackage{amsmath}
\usepackage{amssymb}
\usepackage{graphicx}
\usepackage{mathtools}
\usepackage{braket}
\usepackage{hyperref}
\usepackage{bbold}
\usepackage{calrsfs}
\usepackage{dsfont}
\usepackage{xcolor}
\usepackage{subcaption}

\usepackage{soul}

\makeatletter
\def\@email#1#2{%
 \endgroup
 \patchcmd{\titleblock@produce}
  {\frontmatter@RRAPformat}
  {\frontmatter@RRAPformat{\produce@RRAP{*#1\href{mailto:#2}{#2}}}\frontmatter@RRAPformat}
  {}{}
}%
\makeatother

\usepackage[nohyperlinks, printonlyused, withpage, smaller]{acronym}

\acrodef{DFTB}[DFTB]{density-functional tight binding}
\acrodef{FDTD}[FDTD]{finite-difference time domain}
\acrodef{QEDFT}[QEDFT]{quantum-electrodynamical density-functional theory}
\acrodef{TDDFT}[TDDFT]{time-dependent density-functional theory}
\title{Density-Functional Tight Binding Meets Maxwell: Unraveling the Mysteries of (Strong) Light-Matter Coupling Efficiently}

\author{Dominik Sidler}
  \email{dominik.sidler@psi.ch}
  \affiliation{Laboratory for Materials Simulations, Paul Scherrer Institute, 5232 Villigen PSI, Switzerland }
 \alsoaffiliation{Max Planck Institute for the Structure and Dynamics of Matter and Center for Free-Electron Laser Science, Luruper Chaussee 149, 22761 Hamburg, Germany}
\alsoaffiliation{The Hamburg Center for Ultrafast Imaging, Luruper Chaussee 149, 22761 Hamburg, Germany}%Lines break automatically or can be forced with \\

\author{Carlos M. Bustamante}
  \affiliation{Max Planck Institute for the Structure and Dynamics of Matter and Center for Free-Electron Laser Science, Luruper Chaussee 149, 22761 Hamburg, Germany}
  \alsoaffiliation{The Hamburg Center for Ultrafast Imaging, Luruper Chaussee 149, 22761 Hamburg, Germany}

\author{Franco P. Bonafe}
  \affiliation{Max Planck Institute for the Structure and Dynamics of Matter and Center for Free-Electron Laser Science, Luruper Chaussee 149, 22761 Hamburg, Germany}
  \alsoaffiliation{The Hamburg Center for Ultrafast Imaging, Luruper Chaussee 149, 22761 Hamburg, Germany}

\author{Michael Ruggenthaler}
  \affiliation{Max Planck Institute for the Structure and Dynamics of Matter and Center for Free-Electron Laser Science, Luruper Chaussee 149, 22761 Hamburg, Germany}
  \alsoaffiliation{The Hamburg Center for Ultrafast Imaging, Luruper Chaussee 149, 22761 Hamburg, Germany}

\author{Maxim Sukharev}
  \affiliation{College of Integrative Sciences and Arts, Arizona State University, Mesa, Arizona 85212, USA}

\author{Angel Rubio}
  \email{angel.rubio@mpsd.mpg.de}
  \affiliation{Max Planck Institute for the Structure and Dynamics of Matter and Center for Free-Electron Laser Science, Luruper Chaussee 149, 22761 Hamburg, Germany}
    \alsoaffiliation{The Hamburg Center for Ultrafast Imaging, Luruper Chaussee 149, 22761 Hamburg, Germany}
  \alsoaffiliation{Center for Computational Quantum Physics, Flatiron Institute, 162 5th Avenue, New York, NY 10010, USA}
  \alsoaffiliation{Nano-Bio Spectroscopy Group, University of the Basque Country (UPV/EHU), 20018 San Sebasti\'an, Spain}

\begin{document}

%%%%%%%%%%%%%%%%%%%%%%%%%%%%%%%%%%%%%%%%%%%%%%%%%%%%%%%%%%%%%%%%%%
%                            Abstract                            %
%%%%%%%%%%%%%%%%%%%%%%%%%%%%%%%%%%%%%%%%%%%%%%%%%%%%%%%%%%%%%%%%%%
\begin{abstract}
Controlling chemical and material properties through strong light–matter coupling in optical cavities has gained considerable attention over the past decade. However, the underlying mechanisms remain insufficiently understood, and a significant gap persists between experimental observations and theoretical descriptions. This challenge arises from the intrinsically multi-scale nature of the problem, where non-perturbative feedback occurs across different spatial and temporal scales. Collective coupling between a macroscopic ensemble of molecules and a photonic environment, such as a Fabry–Pérot cavity, can strongly influence the microscopic properties of individual molecules, while microscopic details of the ensemble in turn affect the macroscopic coupling. To address this complexity, we present an efficient computational framework that combines density-functional tight binding (\ac{DFTB}) with finite-difference time-domain (\ac{FDTD}) simulations of Maxwell’s equations (\ac{DFTB}+Maxwell). This approach allows for a  self-consistent treatment of both the cavity and the microscopic details of the molecular ensemble. We demonstrate the potential of this method by tackling several open questions. First, we calculate non-perturbatively two-dimensional spectroscopic observables that directly connect to well-established experimental protocols. Second, we provide local, molecule-resolved information within collectively coupled ensembles, which is difficult to obtain experimentally. Third, we show how cavity designs can be optimized to target specific microscopic applications. Finally, we outline future directions to  enhance the predictive power of this framework, including extensions to finite temperature, condensed phases, and correlated quantum effects. The \ac{DFTB}+Maxwell method enables near real-time exploration of realistic chemical parameter spaces on standard computational resources and offers a systematic approach to bridging the gap between experiment and theory.
\end{abstract}

%\date{\today}

\maketitle
 \section{Introduction} 
\vspace*{-3pt}

%   IMPORTANT NOTICE: The use of this template is not mandatory, and it does NOT represent the FINAL format of the article.   %

\begin{figure}
\includegraphics[width=\columnwidth]{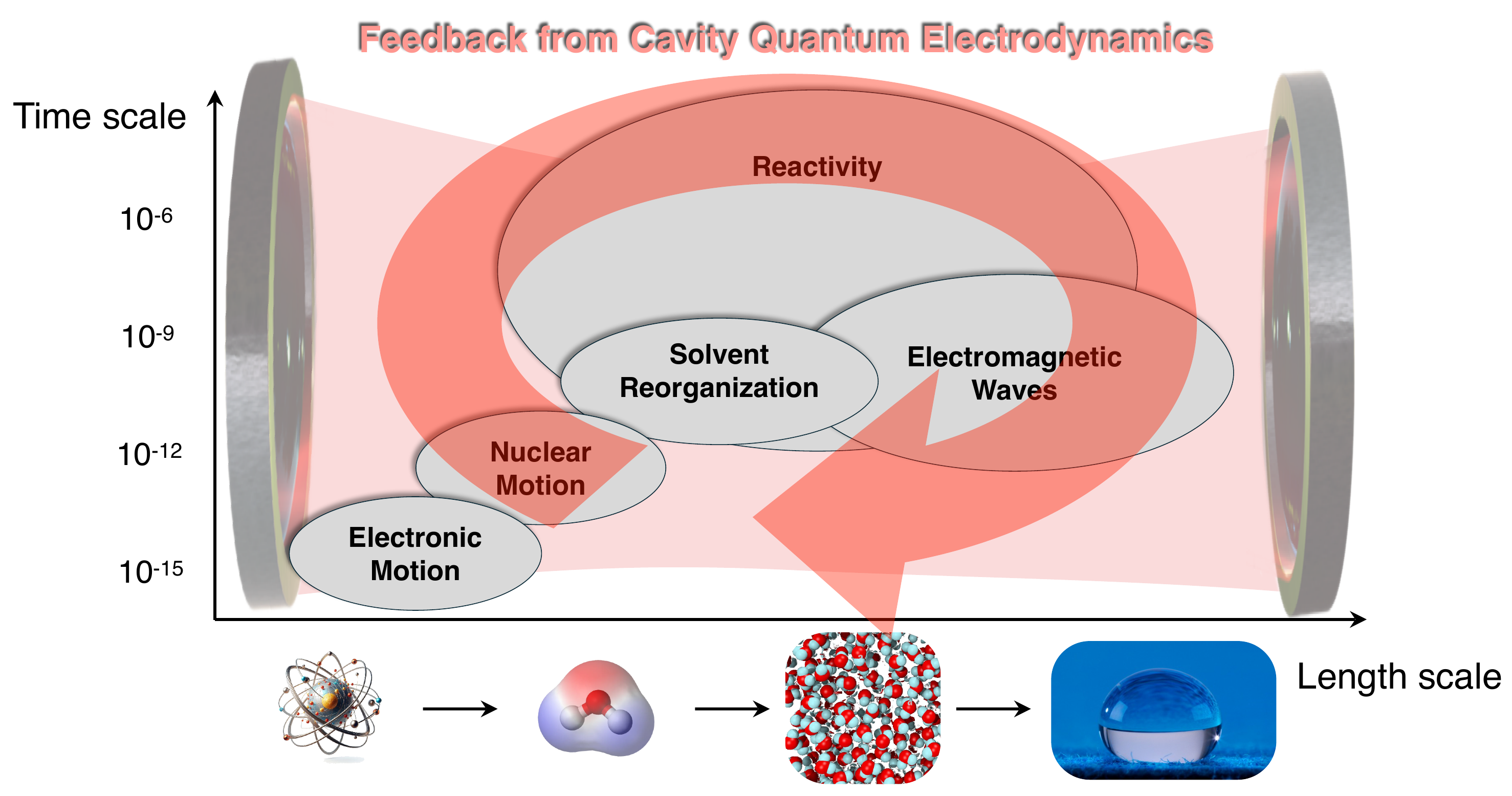}
\caption{In polaritonic chemistry the strong coupling of light and matter in optical cavities introduces feedback effects across different spatiotemporal scales, which are not captured by perturbative or standard computational chemistry methods (illustration taken from Ref.\cite{sidler_solvay}).
\label{fig:feedback}}
\end{figure}

%%%%%%%%%%%
Even a decade after the first experimental evidence showing that strong coupling of molecules in optical cavities can alter chemical properties, such as reaction rates,~\cite{thomas2016a,munkhbat2018a,Hiura2018,lather2019a,thomas2019a,hirai2020a,lather2022a,ahn_modification_2023} energy~\cite{Coles2014,zhong2016a,zhong2017a} and charge~\cite{orgiu2015a,nagarajan2020a} transport, or self-assembly properties,~\cite{hirai2021a,joseph2021a} polaritonic chemistry remains a computationally and theoretically extremely challenging problem. We still lack a detailed mechanistic understanding and microscopic (local) insights on an ab-initio level of theory into chemical ensembles under collective strong coupling.~\cite{ruggenthaler_understanding_2023,sidler_perspective_2022} Collective insights beyond two level systems remain rare exceptions~\cite{sidler_polaritonic_2021,schnappinger_cavity_2023,sidler_unraveling_2024,castagnola2024collective,horak_analytic_2025,sidler2025collectively,castagnola2025realistic}, and many aspects remain open, e.g., the relevance of collective electron correlation effects that possess spin-glass features.~\cite{sidler2025collectively} The reason for the computational complexity is that the collective strong coupling with an optical cavity introduces substantial feedback effects between different spatiotemporal scales, which are not present in traditional theoretical chemistry (see Fig.~\ref{fig:feedback}). Consequently, established approximate methods from quantum chemistry, which have proven their reliability over decades, may no longer capture all the relevant physics. For this reason, different novel ab initio quantum electrodynamics (QED) methods have been developed such as \ac{QEDFT},~\cite{Ruggenthaler2014,Penz2023,Jestadt2019} QED-coupled cluster,~\cite{Haugland2020,Mordovina2020} cavity Hartree-Fock and many more.~\cite{mandal_theoretical_2022, ruggenthaler_understanding_2022} However, in practice those accurate simulation methods are typically restricted to few-molecules strong-coupling simulations in the long-wavelength approximation. Consequently, the spatiotemporal modulations of the transverse electric fields are usually not resolved and the physical complexity of realistic cavity mirrors are discarded. In particular, it is unclear how the interplay of these details affects the interplay between collective and local strong-coupling effects. The Maxwell+\ac{TDDFT} methods that have been developed, both in local dipole approximation \cite{Jestadt2019} and in full minimal coupling \cite{bonafe2025}, are in principle able to consider all these interactions in the classical limit for the photons. However, the microscopic treatment of Maxwell's equations in this formalism renders the application to molecular ensembles impractical due to their multiscale nature. To overcome these spatiotemporal challenges in polaritonic chemistry, we have developed a computational framework that self-consistently couples Maxwell's equations with \ac{DFTB},~\cite{bustamante2025molecular} a highly efficient and scalable electronic-structure method.
This numerical framework allows to investigate non-perturbatively strong light-matter interactions of large polaritonic ensembles in realistic cavities and provides direct access to global and local properties of this multiscale problem. This method is a convenient fast approximation to the full solution of the Maxwell+\ac{TDDFT} equations.

On the ensemble scale, the method can directly compute two-dimensional optical spectra. This experimental technique has emerged as an indispensable tool for probing the complex dynamics of polaritonic systems, where the coherent hybridization of molecular excitations with confined electromagnetic modes creates new pathways for energy transfer, chemical reactivity, and nonlinear optical responses. Unlike conventional spectroscopic techniques, two-dimensional spectroscopy can directly map the coupling networks between polaritons, distinguish homogeneous from inhomogeneous broadening mechanisms~\cite{khalil2003coherent,jonas2003two}, and reveal quantum coherences that arise in cavity-modified chemistry~\cite{dunkelberger2016modified,takele2020multimode}. Such information is essential for the rational design of polaritonic devices, yet remains inaccessible to linear spectroscopy. The interpretation of such experiments, however, demands quantitative theoretical modeling that captures the full complexity of light-matter interaction beyond perturbative approximations~\cite{saurabh2016two}, particularly when molecules are coupled to plasmonic metasurfaces~\cite{sufrin2024probing} or nanostructured cavities, where field confinement, spatial inhomogeneity, and retardation effects fundamentally shape the polaritonic landscape. Current theoretical approaches, predominantly based on simplified cavity models or perturbative expansions, fail to capture the rich physics emerging at plasmonic interfaces, where near-field enhancement, radiative damping, and molecular response are intimately intertwined. The development of computational methods capable of simulating two-dimensional spectroscopic signals from first principles, accounting for the full electromagnetic complexity of metasurfaces and the quantum mechanical nature of molecular systems, thus represents a critical frontier in polaritonic chemistry - one that could bridge the gap between experimental observations and mechanistic understanding of cavity-controlled chemical processes.

On the local scale, the numerical framework allows to investigate how the collective strong coupling affects microscopic details of molecules. Specifically, one can go beyond available results on collectively induced local strong-coupling effects, which commonly assume only a single, spatially homogeneous mode,~\cite{sidler_polaritonic_2021,schnappinger_cavity_2023,sidler_unraveling_2024,castagnola2024collective,horak_analytic_2025,sidler2025collectively,castagnola2025realistic} and investigate the interplay between the spatial inhomogeneity of all the cavity modes and local effects. This allows crucial new insights into the question how collective strong coupling can change local chemical properties. 

Another important aspect of the \ac{DFTB}+Maxwell framework is its computational simplicity, especially for one-dimensional cavities~\cite{bustamante2025molecular}. This allows to emphasize user-friendliness, an aspect that is often overlooked when developing novel simulation tools. Specifically, when interpreting experimental results, 
experimentalists favor simplified effective models over often complex ab-initio light-matter simulation tools. However, to our knowledge, apart from very recent symmetry approaches,~\cite{jayachandranphenomenological} there is no effective model in polaritonic chemistry that can make reliable chemical predictions, beyond (collective) spectral properties (e.g., the collective Rabi splitting). For this reasons, we believe it is vital to develop and provide a user-friendly simulation framework that captures above features and provides reasonable ab-initio predictions on a local computer within short simulation times. Providing such a tool will not only allow to quickly test and explore novel ideas, e.g., modified cavity geometries, losses and frequencies, or different solute-solvent configurations, before performing an experiment, but it will also help scientists to obtain a more intuitive understanding of the feedback between the different scales. We believe the \ac{DFTB}+Maxwell method will be capable to fill this gap.

In the following perspective article, we outline how the self-consistent combination of \ac{DFTB} with a \ac{FDTD} Maxwell solver allows to address all these objectives. We illustrate its current capabilities with three compelling examples and eventually outline its versatility, focusing on different future developments that provide access to unexplored regimes of non-perturbative strong light-matter interaction. Overall, our goal is to introduce a simple but holistic simulation tool for the qualitative and quantitative design of polaritonic experiments and beyond.

\vspace*{-13pt}

\section{Theory and Computation} 
\vspace*{-4pt}

\begin{figure}
\includegraphics[width=\columnwidth]{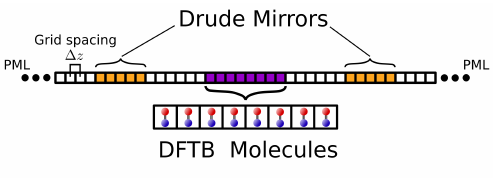}
\caption{Illustration of the one-dimensional grid representation to solve the Maxwell equations for the propagating fields. Notice that the computational approach allows for a mixed representation of polarizable matter. For example, mirrors can be represented by the Drude model, whereas molecules can be described on a \ac{DFTB} level of theory in three dimensions.
\label{fig:grid}}
\end{figure}

The coupling of the \ac{DFTB} Hamiltonian to an arbitrary external electric $\boldsymbol{E}$ follows Maxwell's electromagnetic field equations. As a result, the time-evolution of the electric $\boldsymbol{E}$ and magnetic fields $\boldsymbol{B}$ can be determined  by solving, 
\begin{equation}
    \frac{\partial \boldsymbol{B}(\boldsymbol{r},t)}{\partial t}=-\nabla\times \boldsymbol{E}(\boldsymbol{r},t)\label{eq:maxwell1}
\end{equation}
\begin{equation}
    \frac{\partial \boldsymbol{E}(\boldsymbol{r},t)}{\partial t}= c_0^2\nabla\times \boldsymbol{B}(\boldsymbol{r},t)-\frac{1}{\epsilon_0}\frac{\partial \boldsymbol{P}(\boldsymbol{r},t)}{\partial t},\label{eq:maxwell2}
\end{equation}
discretized on a real-space grid (see Fig.~\ref{fig:grid}). The vacuum speed of light is given by $c_0$ and the vacuum permittivity by $\epsilon_0$. The coupling of the fields to the matter is due to the current density arising from the time-derivative of the polarization $\boldsymbol{P}$

%In our implementation, the time-dependent polarization can be calculated on different levels of theory. For example, the local polarization of metallic mirrors can be represented by the continuum model of Drude as follows,
%\begin{equation}
%    \frac{\partial^2 \boldsymbol{P}(\boldsymbol{r},t)}{\partial t^2}+\gamma\frac{\partial \boldsymbol{P}(\boldsymbol{r},t)}{\partial t}=\epsilon_0\\\omega_p^2 \boldsymbol{E}(\boldsymbol{r},t)
%\end{equation}

In our implementation, the time-dependent polarization can be calculated on different levels of theory. For various metallo-dielectric interfaces such as mirrors, we have implemented several approaches. The simplest representation employs the continuum Drude model for free electrons, while more accurate treatments of material dispersion utilize Drude-Lorentz parametrizations that include inter-band transitions alongside the free-electron response~\cite{rakic1998optical}. We have also implemented modified Lorentz models that can account for additional resonances and damping mechanisms beyond the standard formulations~\cite{deinega2011effective}. For scenarios requiring the description of nonlinear optical phenomena, we employ a semiclassical hydrodynamics model for conduction electrons that provides quantitative treatment of nonlinear effects in metals while maintaining computational efficiency within the Maxwell solver framework~\cite{clark2024harnessing}.

Alternatively, for a more accurate microscopic description of matter, molecular polarizations can be determined by placing single molecules on the Maxwell grid, whose response to the transverse fields is determined from \ac{DFTB} calculations. 
Notice that in our implementation the transverse fields are modified (locally) by the coupling to the matter, thus also affecting the coupling of matter to light at later time-steps. This allows to investigate microscopically, the feedback effects between light and matter.

The \ac{DFTB} Hamiltonian is derived from Kohn-Sham density functional theory by expansion of the energy functional around a reference density $\rho_0$, which is truncated at a certain order (for this study, second order) around the density fluctuations~\cite{Elstner1998,Frauenheim2002}. Then, the molecular orbitals $\psi_i=\sum_\nu C_{i\nu}\phi_\nu$ are represented as a linear combination of atomic orbitals (LCAO) and a minimal basis set of pseudo-atomic orbitals $\{ \phi_\nu \}$ is used. After applying the two-center approximation on the Hamiltonian, and writing the density as a superposition of atomic contributions, the Hamiltonian matrix elements can be written as follows:
\begin{equation}
    \begin{aligned}
    H_{\mu\nu}(t) &= H^{0}_{\mu\nu}+\frac{1}{2} S_{\mu\nu}\sum_K (\gamma_{IK} + \gamma_{JK})\Delta q_K(t),\  \mu \in I, \ \nu \in J.
\end{aligned}
\label{eq:dftbgs}
\end{equation}
Here the sub-indices $I,J$ indicate different atoms, and $q_K$ are the time-dependent Mulliken charges, defined as $q_K = \mathrm{Tr}_K[\rho S]$, being $\rho$ the reduced one-electron density matrix, $S_{\mu\nu}(\boldsymbol{R}) = \langle \phi_\mu|\phi_\nu \rangle$ the overlap matrix elements and $\mathrm{Tr}_A[\cdot ]$ the partial trace over the orbitals centered on atom $A$.
%The \ac{DFTB} Hamiltonian coupled to additional external electric fields $\boldsymbol{E}$, determined by the Maxwell equations, can be written as~\cite{Elstner1998,Frauenheim2002}
%\begin{equation}
%    \begin{aligned}
%    H_{\mu\nu}(t) &= H^{0}_{\mu\nu}+\frac{1}{2} S_{\mu\nu}\sum_K (\gamma_{IK} + \gamma_{JK})\Delta q_K(t) \\
%    &- \frac{1}{2} S_{\mu\nu}(\boldsymbol{\mu}_{I}(\Delta q_I) + \boldsymbol{\mu}_{J}(\Delta q_J)) \cdot \mathbf{E}(\mathbf{r}, t),\ \mu \in I, \ \nu \in J.
%\end{aligned}
%\end{equation}
%Here the sub-indices $I,J$ indicate different atoms, and $q_K$ are the time-dependent Mulliken charges, defined as $q_K = \mathrm{Tr}_K[\rho S]$, being $\rho$ the reduced one-electron density matrix, $S$ the overlap matrix and $\mathrm{Tr}_A[\cdot ]$ the partial trace over the orbitals centered on atom $A$. $\boldsymbol{\mu}_I$ is the dipole moment contribution from atom $I$.
%The molecular orbitals $\psi_i=\sum_\nu C_{i\nu}\phi_\nu$ are represented by a linear-combination-of-atomic-orbitals ansatz and a minimal, localized, non-orthogonal basis set $\{ \phi_\nu \}$.
The non-self-consistent tight-binding Hamiltonian $H_0$, defined as
\begin{equation}
    H_{\mu\nu}^0(\boldsymbol{R})=\langle \phi_\mu|-\frac{1}{2}\nabla^2+v_{\rm eff}[\rho_0]|\phi_\nu \rangle,
\end{equation}
are explicitly pre-computed from DFT as a function of the ineratomic distance $H_{\mu\nu}^0(\boldsymbol{R})$. Together with the overlap matrix elements $S_{\mu\nu}(\boldsymbol{R})$ these quantities are available in Slater-Koster tables, which are used and interpolated for the needed interatomic distance in the calculations.
%The beauty of the \ac{DFTB} approach is that the matrix elements $H_{\mu\nu}^0(\boldsymbol{R}),S_{\mu\nu}(\boldsymbol{R}),\gamma_{IJ}(\boldsymbol{R})$ are precomputed and tabulated for the minimal basis $\phi_\nu$ on \textit{ab initio} calculations.
This allows for a rapid computational evaluation of the eigenvalue problem for the molecular orbitals
\begin{equation}
    \sum_\mu C_{\mu i}( H_{\mu\nu}-\epsilon_i S_{\mu\nu})=0,\label{eq:pseudo}
    \end{equation}
as well as for the update of the time-dependent quantities when doing a time-propagation (see below). Notice, however, that the atomic charge fluctuations $\Delta q_I=q_I(\boldsymbol{C})-Z_I$ depend on the orbital coefficients $C_{\mu\nu}$ with $Z_I$ indicating the charge of a neutral atom of species $I$. Consequently, Eq.~\eqref{eq:pseudo} must be solved self-consistently.

Based on this formalism a real-time time-dependent \ac{DFTB} (rt-TD-DFTB) formulation can be derived from the Runge-Gross theorem~\cite{Niehaus2009,Bonafe2020}. When an additional external electric fields $\boldsymbol{E}$ (for example, given by Maxwell equations) is added, the Hamiltonian matrix can be written as~\cite{Bonafe2020}
\begin{equation}
    \begin{aligned}
    H_{\mu\nu}(t) &= H^{0}_{\mu\nu}+\frac{1}{2} S_{\mu\nu}\sum_K (\gamma_{IK} + \gamma_{JK})\Delta q_K(t)\\
    &- \frac{1}{2} S_{\mu\nu}(\Delta q_I \boldsymbol{R}_I + \Delta q_J \boldsymbol{R}_J) \cdot \boldsymbol{E}(\boldsymbol{R}_{\mathrm{mol}}, t)
\end{aligned}
\label{eq:dftbtd}
\end{equation}
Here the external potential is applied under the electric dipole approximation, where $\boldsymbol{R}_I$ is the position of atom $I$. When coupling to an external field calculated from a Maxwell simulation simultaneously, the value of the external field at the position of the molecule $\boldsymbol{R}_{\mathrm{mol}}$ is considered~\cite{bustamante2025molecular}.

The time-propagation is given in terms of the one-electron density-matrix operator $\rho_{\mu\nu}=\sum_{i\in {\rm occ}}\tilde{C}_{i\mu}\tilde{C}_{i\nu}^*$, and by propagating the well-known Liouville-von Neumann equation for a non-orthogonal basis
\begin{equation}
    \dot{\rho}_{\mu\nu}=-i(S^{-1}\tilde{H}\rho-\rho\tilde{H}^\dagger S^{-1})_{\mu\nu}.
\end{equation}
Here $\tilde{H}_{\mu\nu}={H}_{\mu\nu}-iD_{\mu\nu}$ is a shorthand notation for an effective Hamiltonian that includes both electronic and nuclear dynamics contributions. The non-adiabatic coupling matrix elements defined as $D_{\mu\nu}= \langle \phi_\mu|\dot{\phi}_\nu \rangle$ are considered for Ehrenfest electron-ion dynamics (otherwise they are neglected). In practice these terms are calculated as $D_{\mu\nu}= \sum_\alpha (\partial_\alpha S_{\mu \nu}) \dot{R}_I^\alpha,\, \nu \in I, \alpha=\{x,y,z\}$, where $\dot{R}_I^\alpha$ are the components of the velocity of the nucleus of atom $I$ \cite{Bonafe2020}. The Ehrenfest dynamics at the level of \ac{DFTB} has been successfully applied to the study of light-driven nuclear dynamics in molecules \cite{Bonafe2017a,Hernandez2019}, metallic clusters \cite{Maghrebi2023} and nanoparticles \cite{Bonafe2018a}. Alternatively, Born-Oppenheimer molecular dynamic simulations can also be performed based on the solution of the stationary electronic-structure problem given in Eq.~\eqref{eq:pseudo}, while including the external field in a quasi-static fashion. 

Both the Ehrenfest and Born-Oppenheimer dynamics are implemented in the DFTB+ package~\cite{Hourahine2020}, which we use as a library and coupled to the Maxwell equations by suitable routines of the DFTB+ library that have been implemented for this purpose. Finally, the extended tight binding (xTB) methods are also supported in DFTB+~\cite{Bannwarth2021,Hourahine2020,Hourahine2025}, giving access to simulations for molecules involving almost all elements of the periodic table.

Eventually, the time-dependent molecular polarizations are computed by
\begin{equation}
    \mathbf{P} = N_M \, \langle \hat{\boldsymbol{\mu}} \rangle,\label{eq:polarization}
\end{equation}
and plugged into Eq.~\eqref{eq:maxwell2}. 
Here $N_M$ is a concentration term which gives us the number of molecules represented by the simulated molecule in a single grid point and $\langle \hat{\boldsymbol{\mu}} \rangle = \mathrm{Tr} \left\{ \hat{\rho} \hat{\boldsymbol{\mu}} \right\} $. Notice, our current assignment of molecules to the Maxwell grid, as well as the scaling argument of Eq.~\eqref{eq:polarization} assumes no overlap of the electronic structure between different Maxwell-grid points. Thus any intermolecular interactions remain entirely classical (no exchange and no correlation effects).

\vspace*{-13pt}

\section{Illustrative Applications} 
\vspace*{-4pt}

To demonstrate the capabilities of the \ac{DFTB}+Maxwell approach, we begin by briefly discussing three distinct applications. The first application focuses on global properties, which can be accessed by two-dimensional spectroscopy. Afterwards, we illustrate local molecular changes in optical cavities as well as how to engineer cavity mirrors to reach a desired local effect. In all instances, the Maxwell equations are solved on a grid, however, different quantum descriptions for the matter are applied.

\subsection{Two-dimensional spectroscopy}
\label{subsec:2Dspec}

The Maxwell integrator (see Eqs.~\eqref{eq:maxwell1}\&\eqref{eq:maxwell2}) of \ac{DFTB} plus \ac{FDTD} is under active development, representing a significant advancement in computational electrodynamics for complex light-matter systems. Current implementations support a wide range of dispersive material models, including Drude, Lorentz, and Drude-Lorentz formulations, enabling accurate description of metallic, dielectric, and hybrid structures. The framework handles both linear optical phenomena and nonlinear optical processes including second- and third-harmonic generation, four-wave mixing, and intensity-dependent refractive index changes. These capabilities make the platform particularly suited for studying nanophotonic devices, metamaterials, and strongly coupled light-matter systems.
One of the most exciting recent developments is a novel numerical scheme that enables the calculation of two-dimensional coherent spectra for arbitrary optical systems. Two-dimensional spectroscopy has emerged as an indispensable experimental technique for unraveling complex dynamics in condensed matter systems, providing information about energy transfer pathways, coherent couplings, and many-body interactions that remain hidden in conventional linear spectroscopy. In strongly coupled systems, where light and matter degrees of freedom hybridize to form new quasiparticles, two-dimensional spectroscopy can reveal quantum coherence, distinguish homogeneous from inhomogeneous broadening, and map energy relaxation cascades with unprecedented detail.
Despite its experimental success, theoretical modeling of two-dimensional spectra has remained challenging. The majority of existing theoretical approaches rely on perturbative expansions, phenomenological models, or simplified system-bath interactions that may not capture the full complexity of realistic systems. More critically, truly self-consistent simulations that solve Maxwell's equations simultaneously with quantum mechanical equations of motion have been conspicuously absent from the literature. This gap exists not due to lack of interest, but rather because of a fundamental technical challenge: in \ac{FDTD}-based approaches where electromagnetic fields and material polarizations are evolved self-consistently, the time-domain electric and magnetic fields inherently contain contributions from all orders of the optical response – linear absorption, nonlinear mixing, scattered fields, and cavity feedback effects are all superimposed in the raw simulation data. This mixing makes it seemingly impossible to isolate the specific third-order nonlinear signals that constitute the two-dimensional spectroscopic response, as traditional post-processing techniques cannot disentangle these contributions after they have been combined in the time-domain propagation.

\begin{figure*}
\includegraphics[width=\textwidth]{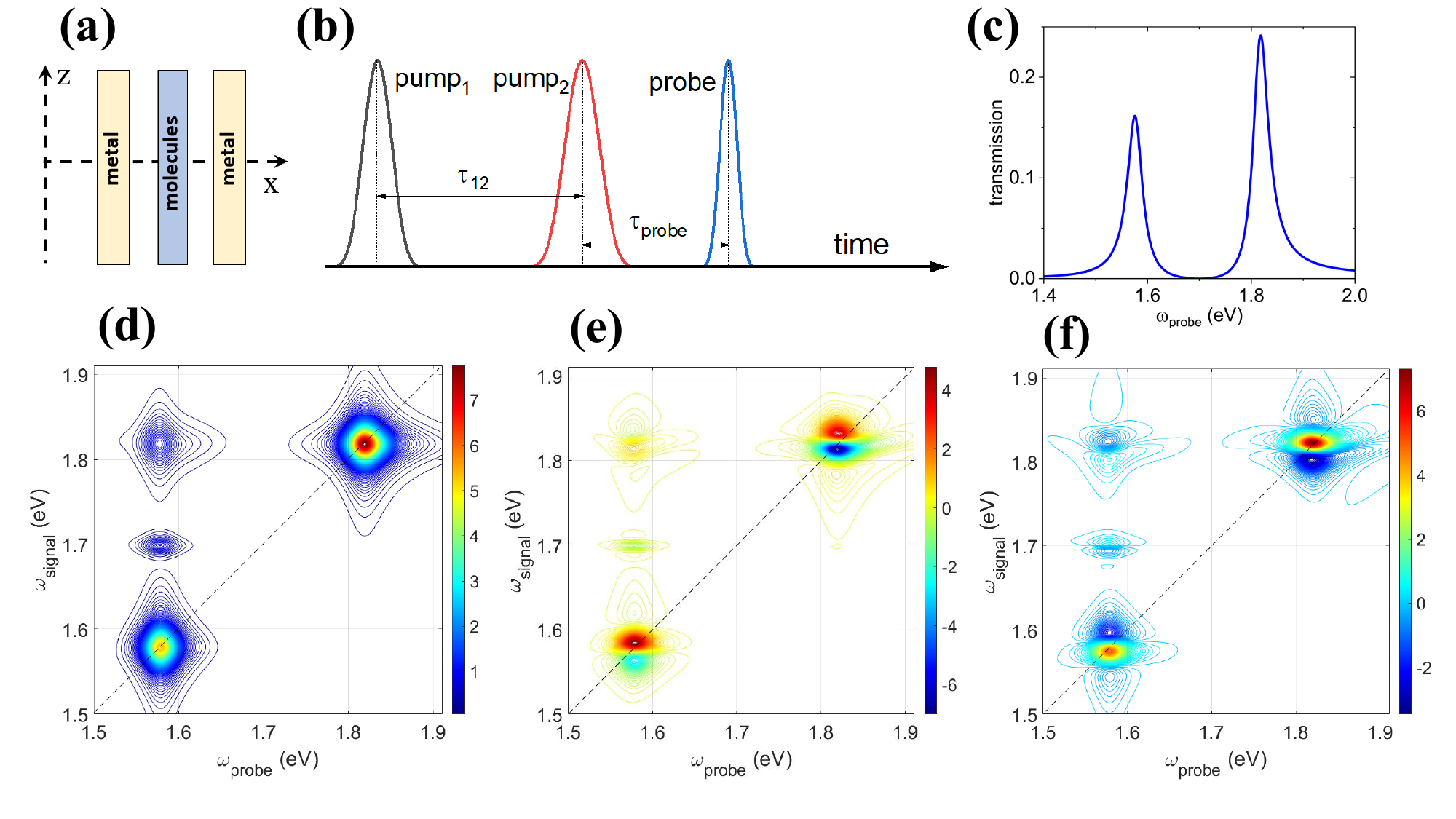}
\caption{Two-dimensional spectroscopy simulations: Panel (a) shows a schematic setup. Panel (b) illustrates the two pumps + probe sequence. Panel (c) shows the linear transmission spectrum indicating lower and upper polaritons. Panels (d) – (f) show the resulting two-dimensional signal as a function of the probe and signal frequencies. Absolute value of the signal is shown in panel (d), panels (e) and (f) show its real and imaginary parts, correspondingly.
\label{fig:2dspec}}
\end{figure*}

The new numerical technique we are developing relies on the propagation of several coupled Maxwell-Schrödinger equations in parallel. Fig.~\ref{fig:2dspec} illustrates the method using a numerical experiment with two pumps and a probe. The signal of interest is the third-order nonlinear process, which is obtained by propagating equations: (1) with all three pulses; (2) with pump 1 only; (3) with pump 2 only. The resulting signal is obtained by subtracting the electromagnetic fields obtained in (2) and (3) from the fields simulated in (1). All calculations are performed in the time domain. A representative example is shown in Fig.~\ref{fig:2dspec}a, comprising two metal slabs sandwiching a molecular layer consisting of two-level emitters treated using standard Bloch equations. The parameters of the system are chosen such that the linear transmission exhibits two well-resolved polariton states shown in Fig.~\ref{fig:2dspec}c. The results of the two-pump/probe sequence are presented in Fig.~\ref{fig:2dspec}d–\ref{fig:2dspec}f. The two-dimensional spectrum reveals the rich nonlinear optical response of the strongly coupled cavity-polariton system through several distinct spectral features. The prominent diagonal peaks at (1.575, 1.575) eV and (1.817, 1.817) eV correspond to lower polariton (LP) and upper polariton (UP) self-interactions, respectively, confirming efficient excitation of both polariton branches with a Rabi splitting of 242 meV. A clear cross-peak at (LP, UP) demonstrates coherent coupling between LP and UP states, while the absence of the corresponding (UP, LP) peak indicates asymmetric relaxation favoring energy flow from upper to lower polariton branches. Most intriguingly, a narrow peak appears at (LP, 1.698) eV, where the signal frequency matches the bare emitter transition, suggesting the important role played by the dark states. 

The computational framework demonstrated here for two-dimensional spectroscopy of strongly coupled cavity-emitter systems represents just one facet of the broader capabilities enabled by interfacing the Maxwell integrator with \ac{DFTB} electronic-structure methods. By combining the non-perturbative propagation of the full Maxwell-matter dynamics with \ac{DFTB}'s efficient treatment of molecular electronic structure, our approach surpasses the limitations of current methodologies that rely on perturbative expansions~\cite{mukamel1995principles} - limitations that become particularly severe in the strong coupling regime where light-matter hybridization fundamentally alters both optical and molecular properties~\cite{reitz2025nonlinear}. The self-consistent treatment of both the electromagnetic cavity modes and molecular degrees of freedom within a unified computational framework eliminates the need for artificial partitioning between "system" and "bath" that underlies most contemporary approaches, instead capturing the full complexity of polaritonic states and their nonlinear optical response from first principles. This capability to perform quantitative simulations of multi-dimensional spectroscopy without invoking rotating wave approximations, Markov assumptions, or perturbative field-matter coupling represents a significant advance toward predictive modeling of cavity-modified chemistry and opens new avenues for designing and optimizing polaritonic devices, where strong coupling effects dominate the system dynamics.

\subsection{Collective Strong Coupling in Optical Cavities}
\label{subsec:collective}

Experimentally, collective strong coupling is manifested as a larger separation of the polaritonic peaks in the transmission spectrum (Rabi splitting) when the concentration of molecules inside the cavity is increased. To capture this phenomenon, we solve the Maxwell equations on a one-dimensional grid, while different numbers of molecules are described using \ac{DFTB}. The molecules are always resolved in three dimensions, providing a realistic representation of molecular structures. Allocating two sections of the one-dimensional grid to represent the mirrors, using the Drude model, it becomes possible to mimic a Fabry-P\'{e}rot cavity.

\begin{figure}
\includegraphics[scale=0.5]{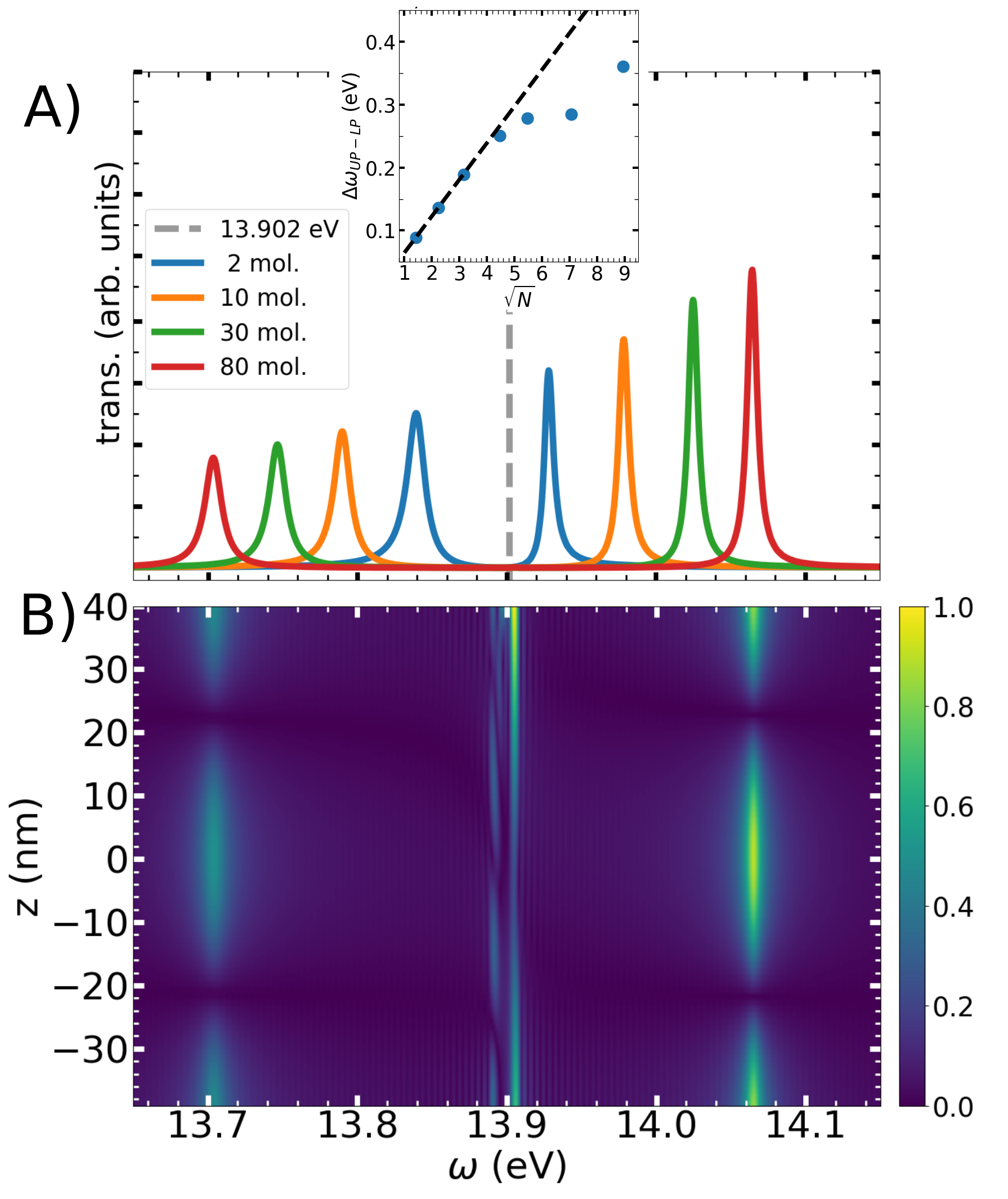}
\caption{A) Transmission spectra obtained after exciting the cavity and different number of simulated molecules. The gray dashed line indicates the first electronic excitation energy of N$_2$. Inset: Rabi splitting as a function of the square root of the number of molecules. B) Fourier-transformed dipole moments of the simulated molecules as a function of their position in the cavity, for the system comprising 80 molecules. The color scale has been normalized with the maximum amplitude of the dipole moment.
\label{fig:spec_dip}}
\end{figure}

As a conceptual example, we work with nitrogen molecules ($\rm N_2$), which possess a first electronic transition at 13.902 eV, according to \ac{DFTB}. To construct a cavity that resonates with this transition, we set two Drude mirrors separated by 123 nm. The mirror thickness was chosen 20 nm  and the Drude parameters were set to $\Omega_p=$ 34 eV and $\gamma=$ 0.181 eV. Under these conditions, the third mode of the cavity resonates with the electronic transition. Figure~\ref{fig:spec_dip}A shows the different transmission spectra obtained by explicitly varying the number of molecules simulated. As we can see, increasing the number of molecules leads to a larger Rabi splitting. The inset in Figure \ref{fig:spec_dip}A shows the dependence of the Rabi splitting on the square root of the number of molecules $N$. It can be seen that the typical linear scaling with $\sqrt{N}$ breaks down for a certain amount of molecules. This is because the linear behavior is obtained assuming that all the molecules experience the same electric field, while in these simulations each molecule is subject to different field amplitudes depending on its position. For a small number of molecules located around the center of the cavity, the space occupied by them is less than half the wavelength of the excitation field, and therefore the field is almost constant and the linear dependence of the Rabi splitting with respect to $\sqrt N$ is reproduced. However, for a larger molecular ensemble, their spatial extent becomes comparable to the wavelength of the resonant cavity mode, breaking the $\sqrt{N}$ behavior. We thus see that the spatial distribution can have a profound impact on these relations.

To shed more light on the consequences of the spatial distribution of the cavity modes, we next analyze individual molecular properties. By Fourier transforming the time-dependent dipole moment, we can obtain information about the spectral response of each molecule. In Fig.~\ref{fig:spec_dip}B, the Fourier transform of the dipole moment for the case of 80 molecules is shown as a function of the position of the molecules. We can observe that the polaritonic peak intensities (at 13.7 eV and 14.06 eV) exhibit a spatial distribution that matches the expected profile of the cavity's third mode, with nodes near one-third and two-thirds of the mirror separation distance ($z \approx \pm 22 \; \rm nm$ in the plot), evidencing local effects as a result of the realistic description of the molecular distribution in the cavity. In fact, there is experimental evidence of the position-dependent nature of the Rabi splitting when a J-aggregate layer is displaced inside the cavity, following a mode-like profile~\cite{Wang2014}.

Figure~\ref{fig:spec_dip}B also reveals more features than are present in the transmission spectra (Figure \ref{fig:spec_dip}A). The signals obtained above and below 13.9 eV correspond to dark states created by the electronic excitation of the molecules and to the off-resonant interaction between the molecules and higher modes, respectively. These results demonstrate the ability of our implementation to access local information that is lacking in the transmission spectrum.

%Next we look at an ensemble of molecules under collective strong coupling in an explicit optical cavity. The Maxwell's equations are solved on a 2D grid, whereas the DFTB molecular description is consistently resolved in 3D, thereby providing a realistic representation of molecular structures.
%First,  201 nitrogen molecules were aligned equidistantly at the center of a Fabry-Perot cavity, modeled explicitly by two Drude mirrors (see Fig. \ref{fig:2dcavity}), CHARLY:GIVE DETAILS E.G. DISTANCE TUNED ON RESONANCE WITH WHAT. HIGHLIGHT THAT WE CAN LOOK AT COLLECTIVE STRON COUPULING EFFECTS. Driving the cavity gives access to spatially resolved spectral and molecular properties. Having access to spatially resolved physical properties, is particularly, relevant for the rational design of complex cavity geometries, to specifically target microscopic properties of matter.  Moreover, it is  also useful to develop  an intuitive microscopic understanding of  polaritonic processes in general.

\subsection{Automatized Cavity Design}
\label{subsec:cavitydesign}
  Inverse design is a powerful methodology for engineering optical devices that fulfill predefined characteristics. Among the many existing approaches, density-based topology optimization methods are the most flexible and versatile~\cite{molesky2018inverse, hammond2022high}. We have included this tool in our implementation to design optical cavities in two dimensions, optimized for strong coupling with a selected molecule and to enhance the interaction between groups of the same molecule at different positions and scales.
  
\begin{figure*}
\centering
\includegraphics[scale=1.2]{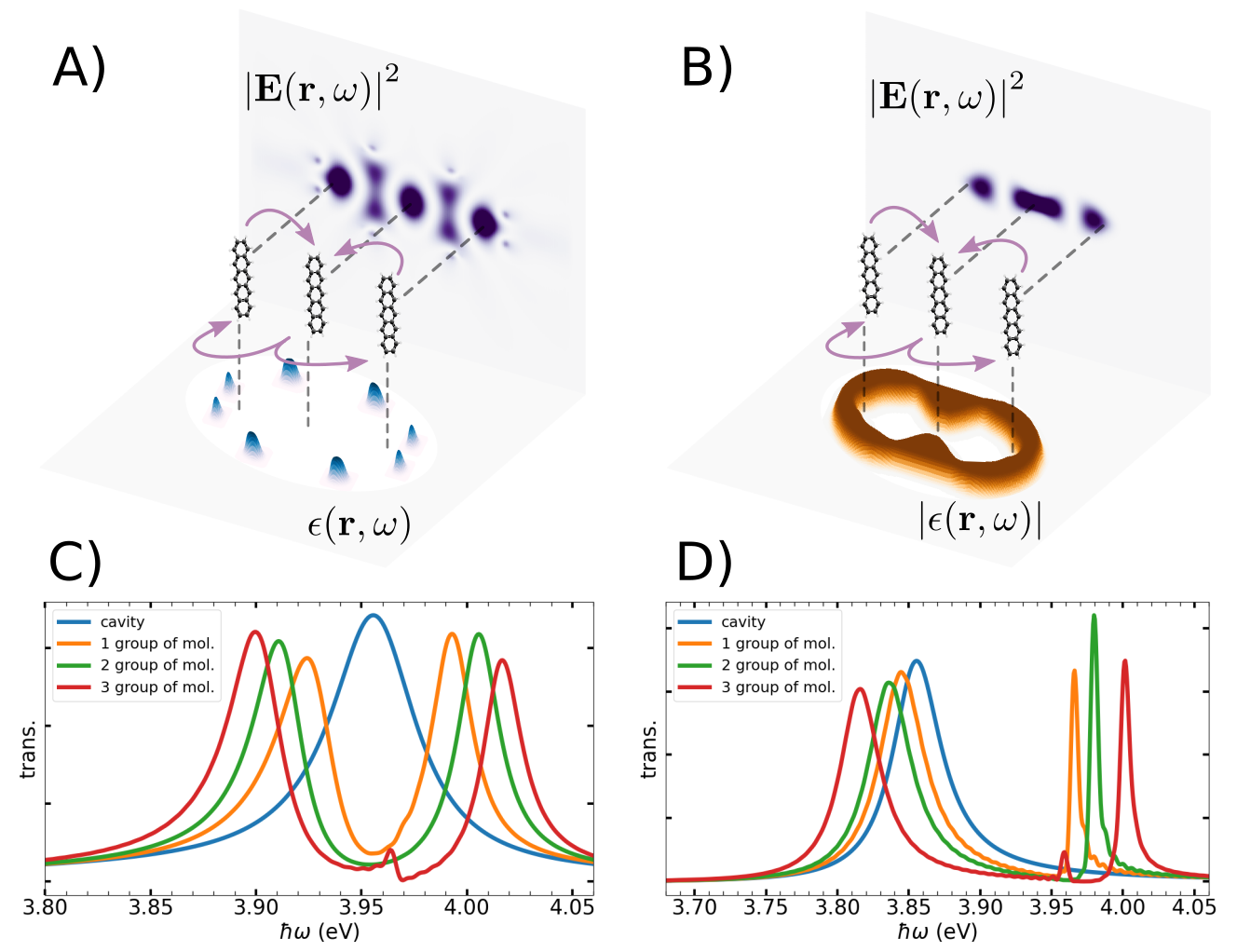}
\caption{A) and B) present two examples of cavities designed in 2D using two different materials, silicon and gold respectively. The designed cavities were extended into the third dimension for visualization purposes.
The obtained designs aim to maximize $\left| \mathbf{E}(\omega) \right|^2$ at two target points located at (-300 nm, 0 nm) and (300 nm, 0 nm), when a point-like current source is placed at (0 nm, 0 nm). The design frequency was 3.97 eV, corresponding to the first electronic transition of the pentacene molecule. The resulting $\left| \mathbf{E}(\omega) \right|^2$  distribution in the simulation box, for the respective cavities, is also shown in these panels. C) and D) display the transmission spectra obtained from DFTB+FDTD simulations after exciting the system with a short point-like pulse, positioned at (0 nm, 0 nm) and tuned to  $\omega=$ 4.0 eV, within the silicon and gold cavities, respectively. The spectra compare the empty cavity (blue curve) with cavities containing different numbers of groups (37 molecules each), located at one  target point (orange curve), two target points (green curve), and two target points and source point (red curve).
\label{fig:invdesign}} 
\end{figure*}

  The algorithm solves the Maxwell equations in the frequency domain on a two-dimensional grid at a fixed frequency corresponding to a chosen molecular transition. The optimization seeks to maximize $\left| \mathbf{E}(\omega) \right|^2$ at one or more target points, when a point-like current source is located at a specified position in the simulation box. To this end, we calculate the gradient $d\left| \mathbf{E}(\omega) \right|^2/d\varepsilon(\mathbf{r}, \omega)$, where $\varepsilon(\mathbf{r}, \omega)$ is the corresponding permittivity at the frequency $\omega$ and in the grid point with coordinate $\bf r$. Using this gradient, the permittivity at each grid point within a selected region is updated continuously in each optimization step. A binary projection is applied so that the permittivity takes values corresponding to either the chosen design material or vacuum. The obtained designs are tested afterwards in our \ac{DFTB}+Maxwell implementation including the molecules explicitly. For example, Fig.~\ref{fig:invdesign}A and B show two cavities made of silicon and gold, respectively, designed to be in resonance with pentacene molecules (electronic transition at 3.97 eV), and to couple three groups of molecules separated by 300 nm. The designs aims to maximize $\left| \mathbf{E}(\omega) \right|^2$ at two target points located at (-300 nm, 0 nm) and (300 nm, 0 nm), when a point-like current source is placed at (0 nm, 0 nm). In the same figure, a profile of the resulting $\left| \mathbf{E}(\mathbf{r}, \omega) \right|^2$ in the whole simulation box is also plotted, demonstrating the desired increase at the selected points as well as at the source point. To test each cavity, we conducted three simulations including one, two, and three groups of 37 pentacene molecules, respectively, located at one target point, both target points, and both target points along with the source point. In all cases, the system is excited with a short point-like pulse, positioned at the source point and tuned at $\omega=$ 4.0 eV. The signal transmitted out of the cavity is utilized to obtain the transmission spectra presented in Fig.~\ref{fig:invdesign}C and D, for the silicon and gold cavities, respectively. In both cases, we observe that the presence of just one group of molecules already yields a Rabi splitting, indicating that we are in the strong-coupling regime. As we increase the number of groups of molecules, the separation between the polariton signals increases. This demonstrates that, despite the large distance between the target points, the different groups of molecules are interacting through the cavity mode and form a polaritonic ensemble.

  Even though the presented example focused on the optimization of $\left| \mathbf{E}(\omega) \right|^2$ through modification of the permittivity $\varepsilon(\mathbf{r}, \omega)$, our implementation can be extended to the optimization of other field magnitudes, as well as those related with the magnetic field $\mathbf{B}(\mathbf{r}, \omega)$. Some of these cases will require the magnetic permeability ($\mu(\mathbf{r}, \omega)$) as an additional optimization variable, which can be included straightforwardly.

  %an inverse design algorithm was applied to do design cavity mirrors made of ???, which maximize the (CHARLY: BE MORE PRECISE) interactions between two or three Benzene molecules macroscopic distances (i.e., beyond London dispersion effects) (see Fig. \ref{fig:invdesign}). From the inverse design simulations one notices that depending on the chosen chemical setup and purpose, very complex cavity geometries can be demanded. Those geometries are only computationally accessible if detailed molecular (electronic structure properties) are resolved in combination with the feedback of the macroscopic fields, shaped by the mirrors and surrounding molecules. Notice, again that the beauty of the DFT+Maxwell approach is that above examples  could all be solved on a single core  (CHARLY: GIVE DETAILS). Thus no dedicated HPC infrastructure is necessary for spatio-time-resolved predictions. This makes our chosen computational approach particularly appealing to experimentalists and non-expert users.

\section{Outlook}

While the \ac{DFTB}+Maxwell framework is already able to address various different problems of polaritonic chemistry and strong light-matter coupling, there are several directions that are actively developed and will enhance the capabilities of this theoretical toolkit even further. In the following we briefly sketch the main directions:

\begin{enumerate}
    \item \textbf{Thermal Equilibrium:}
    In the context of thermal effects, the self-consistently coupled \ac{DFTB}+Maxwell framework represents a more general problem than usually encountered in standard ab-initio Molecular Dynamics (MD) setups (e.g., Ehrenfest or Born-Oppenheimer MD). Here, not only the nuclear degrees of freedom can be coupled to a thermal bath (e.g., by Langevin equations), but also the field can be thermostated, for example, by explicitly accounting for the black-body spectrum in the transverse fields \cite{gadea2022radiative}. The inclusion of transverse degrees of freedom breaks the usual conservative-field assumption and hence introduces the possibilities of novel effects. Notice that under collective strong-coupling conditions, the definition and characteristics of thermal (non)-equilibrium becomes a non-trivial problem of fundamental theoretical interest,~\cite{sidler_perspective_2022,sidler_numerically_2023,sidler2025collectively} but also practical relevance. For example, modified thermodynamics would alter chemical reaction rates and thus corresponding theoretical considerations~\cite{montillo_vega_theory_2024,ying_resonance_2024,yang_quantum_2021,ke_insights_2023,ke_quantum_2024,lindoy_quantum_2023,Lindoy2022} would require adjustment. For example, from a statistical point of view, the strongly coupled molecular ensemble can no longer be partitioned into weakly interacting sub-entities that can be regarded statistically independent and form the basis for a thermal statistical description of large molecular ensembles.~\cite{sidler_numerically_2023} Moreover, on resonance, theoretical models show that cavity-mediated inter-molecular electron correlations, i.e., London-dispersion forces, are substantially modified, which can lead to spin-glass-like effects (see point 4 below) that indicate the violation of standard fluctuation-dissipation relations.~\cite{sidler2025collectively} 
    The tackle these fundamental theoretical challenges requires efficient and versatile computational approaches. We believe the \ac{DFTB}+Maxwell framework is ideal to investigate the above mentioned issues. 
    \item \textbf{Beyond Gaseous Phase:}
    In our compact theoretical introduction of \ac{DFTB}+Maxwell, we have focused on gas-phase molecular setups. To account for liquid condition, i.e., longitudinal inter-molecular interactions, it is straightforward to replace the \ac{DFTB} molecules placed on the Maxwell grid by explicit molecular ensembles on a \ac{DFTB} level of theory. For example, this would allow to explicitly account for the solvation shells surrounding the solute molecules. Moreover, using the basics of \ac{QEDFT}, which makes \ac{DFTB} a limiting case of \ac{QEDFT}, we can extend density-functional embedding methods to account also for longitudinal and transverse interactions between the molecular ensembles at different (macroscopic) Maxwell grid points.  

   Using such an embedding strategy, it becomes possible to also include local (within one macrosopic grid point) transverse field effects in the \ac{DFTB}+Maxwell setup. However, these single-grid-point local-field effects might not be overly important. The fully quantized Pauli-Fierz theory reveals that transverse local-field effects (from the dipole self-energy term) are expected to be significant mainly on intermediate to large length scales, exceeding typical Coulomb correlation lengths.~\cite{sidler2025collectively} Thus, the explicit modeling of these interactions via the macroscopic Maxwell equations as done in the Maxwell+\ac{DFTB} might be accurate enough. 
    \item \textbf{Collective Quantum / Correlation Effects:}
    
    Choosing the classical Maxwell equations to propagate the macroscopic transverse and longitudinal fields in time, a priori removes any physical effect that may emerge on this scale due to the quantization of the electro-magnetic field. In particular, the restructuring of the continuum of vacuum modes cannot be resolved. However, this is considered a crucial ingredient for going from weak to strong coupling conditions. For example, in weak coupling the vacuum-field fluctuations are responsible for the Casimir forces and the Purcell enhancement of spontaneous emission. Under (collective) strong coupling the relevance of vacuum fluctuations are still under debate. In particular, it remains unclear to what extend and under which conditions the quantum nature of the electromagnetic field is key and surpasses standard thermal fluctuations. Experimental evidence suggest that in particular meeting collective-resonance conditions and (modified) London dispersion forces are crucial ingredients to explain changes of chemical properties under collective strong coupling. For example, modified bulk properties such as self-aggregation or changed conformational equilibrium constants were observed. In addition, rare events (tunneling and chemical reaction rates) could also be altered. Those chemical effects are believed to be a consequence of the quantized nature of the transverse electric fields. In more detail, as recently shown in Ref.~\cite{sidler2025collectively}, those effect could be triggered by the spin-glass-like nature of transverse electron correlations in optical cavities. Moreover, the explicit theoretical mapping onto a spin glass does not only support aforementioned experimental facts theoretically, but it also predicts phase-transitions, heating and hysteresis effects under resonant collective strong coupling.~\cite{sidler2025collectively} All of which have been observed experimentally in various situations.~\cite{patrahau_direct_2024,joseph2021a,sandeep_manipulating_2022,pang2020a,kaur_controlling_2023,kumar_extraordinary_2024,biswas_electronic_2024,jarc_cavity-mediated_2023,zhong_driving_2023,thomas_large_2021} 
    
    In order to include these quantum effects in the \ac{DFTB}+\ac{FDTD} scheme, we can again rely on the fact that \ac{DFTB}+Maxwell is the mean-field limit of \ac{QEDFT}. Thus, by including transverse exchange-correlation functional approximations (on top of the standard longitudinal exchange-correlation functional approximations, which describe the Coulomb interaction, and are already included in \ac{DFTB}+Maxwell) we can efficiently and accurately simulate these novel quantum effects also in this framework. 

    \item \textbf{Photon-field Fluctuations by Semiclassical Methods:} Another research direction that can be explored with \ac{DFTB}+Maxwell methods is the possibility to recover quantum effects of light such as vacuum fluctuations and correlations via multi-trajectory Ehrenfest methods. These approaches are based on propagating a swarm of Ehrenfest trajectories, where the classical photon field coordinates are sampled from a Wigner distribution, and the observables are ensemble averages of the individual trajectories~\cite{Hoffmann2019PRA, Hoffmann2019JCP}. Among the different semiclassical methods that haven been bench-marked, the multi-trajectory Ehrenfest method qualitatively describes the dynamics of electron and photon populations and captures bound photonic states. While these calculations have been done considering a mode-resolved version of the Maxwell equations, they can be adapted to the case of spatially-resolved fields. One of the advantages of \ac{DFTB} simulations for this resides in the large amount of trajectories that are needed, on the order of $10^5$, to achieve converged results. While this number becomes prohibitively expensive for full ab-initio simulations of realistic systems, they become feasible by the use of \ac{DFTB}.  
    \item \textbf{Nuclear Quantum Dynamics:}
    So far our implementation of \ac{DFTB}+Maxwell only allows for classical trajectories of the nuclei. However, for molecules in a dilute gas phase, quantum dynamical effects can start to play a role. For example, the rotational dynamics of apolar diatomic gases can be well approximated by the quantum rigid rotors. In that sense, it would be very interesting to investigate the feedback between classical Maxwell fields and the quantized dynamics of nuclei. At least for dilute gases, our framework should be extendable to incorporate nuclear quantum dynamics for example by replacing the \ac{DFTB} part by multi-configuration time-dependent Hartree propagations~\cite{beck2000multiconfiguration,kato2018time} or hierarchical equations of motion~\cite{tanimura2020numerically,batge2021nonequilibrium}. Clearly, the time-propagation would be restricted to small molecules and short simulation times. Still, collective feedback effects on the quantum dynamics should become accessible for many molecules thanks to the dilute gas limit. In that case, the different molecules can be considered interacting only via classical Maxwell fields. This effectively suppresses the scaling of the quantum dynamical complexity with respect to the number of molecules.

    \item \textbf{Simplicity and User-Friendliness:}
    The gap between theory and experiments in polaritonic chemistry is enhanced due to the vastly different interpretational tools. Since currently most numerical ab-initio QED tools are not easy to use without much prior experience, experimentalist favor simplified models to interpret the observed effects. We therefore think that an efficient and user-friendly numerical toolkit will be central to bridge this gap. That is, we aim to develop a framework that can be easily employed to interpret experimental results on the fly, and even allows to optimize experimental details in-situ. Thanks to the computational speed of the one-dimensional Maxwell solver in combination with \ac{DFTB}, it becomes possible to aim for cloud platforms, which allow users to execute tasks directly from their web browser. This reduces the entry barrier for setting up a computational research environment.~\cite{yakutovich2021aiidalab} In that regard, we will examine the possible integration of \ac{DFTB}+Maxwell into AiiDAlab. AiiDAlab is a web platform that enables computational scientists to package scientific workflows and computational environments and share them with their collaborators and peers. It allows for user-friendly graphical interfaces of key functionalities and is made available through the AiiDAlab app store.~\cite{yakutovich2021aiidalab} Still more complex functionalities, that are only applicable on large high-performance computing facilities, can be developed concurrently.

    \item \textbf{Three-Dimensional Maxwell Grid:}
    Our Maxwell solver incorporates advanced parallelization capabilities that enable simulations of large ensembles of quantum emitters in complex electromagnetic environments. The implementation employs a novel molecular-mapping technique built on top of the conventional three-dimensional domain decomposition methodology, which efficiently balances computational load across parallel processors and allows direct access to collective dynamics by drastically increasing the number of molecules that can be simulated simultaneously~\cite{sukharev2023efficient}. This parallelization framework is being merged with electronic-structure codes, enabling us to perform large-scale simulations of many molecules interacting with complex three-dimensional metasurfaces. The combination of efficient parallel computing with first-principles molecular calculations opens new possibilities for studying collective optical phenomena in realistic nanostructured environments, where both the electromagnetic field distribution and the quantum mechanical response of individual emitters are treated with high fidelity.~\cite{sukharev2025unveiling} 
    
\end{enumerate}

\section{Summary and Conclusion}

This perspective article demonstrates the numerous capabilities and opportunities of a computationally efficient and versatile simulation framework for (strong) light-matter interaction. The \ac{DFTB}+Maxwell framework bridges macroscopic classical field propagation with microscopic molecular resolution on a quantized level of theory, including their respective feedback effects in space and time. Capturing this delicate interplay offers non-perturbative, quantitative insights into (multi-dimensional) spectroscopic observables (see e.g. Sec.~\ref{subsec:2Dspec}) with detailed access to local chemical changes, for example induced by the hybridization of light and matter inside optical cavities (see e.g. Sec.~\ref{subsec:collective}). Eventually, this computational framework can be employed to specifically design the photonic environment on demand (see e.g. Sec.~\ref{subsec:cavitydesign}). Based on these fundamental capabilities, we have highlighted different opportunities to enhance the predictive power of this toolkit, including temperature or collectively correlated quantum effects for coupled light and matter in a non-perturbative fashion. However, all of these effects can lead to an intricate feedback between different scales, and hence common simplified interpretational tools can fail to capture these important details. Therefore, an essential ingredient for the success of this computational framework is its computational simplicity and speed that extends to collective effects of strong light-matter interaction. Fortunately, for many practical cases the one-dimensional representation of the Maxwell fields is sufficient together with a three-dimensional representation of molecular ensembles. In this case, the \ac{DFTB}+Maxwell method allows for an almost real-time exploration of chemical or physical parameters on a desktop machine. We believe our approach provides a great opportunity to be developed into a user-friendly tool suitable for in-situ interpretation of experimental setups and the design of novel industrial applications, without having to rely on over-simplified effective model systems or dedicated  high-performance computing infrastructure.

%%%%%%%%%%%%%%%%%%%%%%%%%%%%%%%%%%%%%%%%%%%%%%%%%%%%%%%%%%%%%%%%%%%%%%%%%%%%%%%%%%
\begin{acknowledgement}
We gratefully acknowledge all other members of our ERC synergy grant team UnMySt for inspiring discussions (Tal Schwartz, Thomas Ebbesen, Abraham Nitzan,  Sharly Fleischer, Cyriaque Genet). We acknowledge support from the Max Planck-New York City Center for Non-Equilibrium Quantum Phenomena. The Flatiron Institute is a division of the Simons Foundation.
This work was made possible by the  European Research Council, Grant No. ERC-2024-SYG-101167294, UnMySt.   
M.S. research is supported by the Office of Naval Research, Grant No. N000142512090.
C. M. Bustamante thanks the Alexander von Humboldt-Stiftung for the financial support from the Humboldt Research Fellowship.
F.P.B. acknowledges financial support from the European Union’s Horizon 2020 research and innovation program under the Marie Sklodowska-Curie Grant Agreement no. 895747 (NanoLightQD).
\end{acknowledgement}

\bibliography{references}
\end{document}